# A new application of Multimodal Biometrics in home and office security system

K.Saravanan[1], C.Saranya[2] and M.Saranya[3]

[1] Asst.Professor/CSE, Erode Sengunthar Engineering College, Erode, India

[2,3] B.E. C.S.E. Student, Erode Sengunthar Engineering College, Erode, India

Email: saravanankumarasamy@gmail.com

*Abstract* - *Biometric door lock security systems are used at those places where you have important information and stuffs. In that kind of places multibiometric electronic door lock security systems that are based on finger print and iris recognization. Multibiometric door lock security systems are used to prevent the door related burglaries such as break ins occurred in different forms so this is the best method to prevent this type of happenings. Unlike keyed locks, there is no need to take the keys with you when you go out without necessary of worrying about losing keys. This paper proposes a multimodal biometrics door lock system with iris and fingerprint as a computer application for automatically identifying or verifying a person from fingerprint iris recognition system. The first stage is identification and second one is verifying that whether he is a genuine user or imposter. During second stage system compares the input set with all available stored set in database. This comparison gives a ranked list of matches. Based on the rank retrieved an alarm is activated automatically when any unauthorized person tries to open the door. So this kind of multimodal biometrics will provide a highly secured and authenticated access.*

## I. INTRODUCTION

Door-access control is a physical security system that assures the security of a room or building by means of limiting the access to that room or building to specific people and by keeping records of such accesses. It utilizes an individual-authentication method in order to limit access to specific people. The most widespread authentication method for such systems is based on smart cards. Such a system limits room access to only those people who hold an allocated smart card. However, in the case of smartcard systems, on top of the difficulty in preventing another person from attaining and using a legitimate person's card, there is the inconvenience of processing lost cards. In the meantime, accompanying the continuing development of fingerprints as the main biometrics method for individual authentication, the practical application of door-access-control systems utilizing biometric data has begun. Biometrics authentication uses information specific to a person's body in order to assure a high level of security that makes it difficult for a stranger to impersonate that person. Although there are several types of biometrics authentication methods, the one — called multibiometrics authentication — presented here is the most suitable method for controlling door access by a large number of people. This paper describes the features of the developed multibiometrics authentication method and presents two applications of this authentication method to door-access control. Biometrics is the science of identifying or verifying an individual based on the physiological or behavioral characteristics like face, fingerprint, iris, signature, voice, retina, handwriting, and so forth. Unimodal biometric system uses a single source of biometric information for personal identification. In spite of numerous advantages of biometrics-based personal authentication systems over traditional security systems based on token or knowledge, they are vulnerable to attacks that can decrease their security considerably, which include many problems: noisy sensor data, lack of individuality and non-universality The purpose of multimodal biometric is to overcome the limitations of the unimodal, while a better performance can be obtained by combining the evidences presented by multiple traits of fingerprint, iris, face and so on. To deal with the problem that how multibiometrics systems outperform the traditional unimodal biometric system, many fusion solution have been discussed. By using more then one means of biometric identification, the multimodal biometric identifier can retain high threshold recognition settings. The system administrator can then decide the level of security he/she requires. With this methodology, the probability of accepting an impostor is greatly reduced.

## III. FINGER PRINT RECOGNITION SYSTEM

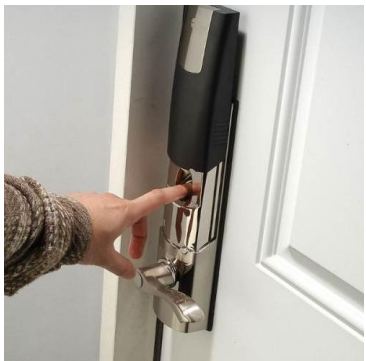
Fig 1. Fingerprint recognition System

The first step is the capturing of a fingerprint image. This would normally be done using a fingerprint scanner or Reader which was attached with the dorrlock.. The fingerprint image is passed to the recognition software for recognition number of steps such as normalizing the fingerprint image and then creating a "template" of "print" to be compared to those in the database. The match can either be a true match who would lead to investigative action or it might be a „false positive which means the recognition algorithm made a mistake and the alarm would be cancelled. Each element of the system can be located at different locations within a network, making it easy for a single operator to respond to a variety of systems.

*A. Fingerprint Verification*

This paper introduces a prototype automatic identity authentication system that is capable of authenticating the identity of an individual, using fingerprints. The main components of the AFIS are
   1. Fingerprint database
   2. Fingerprint features database
   3. Enrollment Module
   4. Authentication module.

The fingerprint database stores the fingerprint images. For this purpose. The features extracted from these fingerprints are stored in the features database along with the person ID[3]. The objective of the enrollment module is to admit a person using his/her ID and fingerprints into a fingerprints database after the process of feature extraction. These features form a template that is used to determine or verify the identity of the subject, formulating the process of authentication. The component of the AFIS used for authentication is referred to as the authentication module. Figure 3 illustrates the different steps involved in the development of the AFIS. The details of these steps are given in the following subsections.

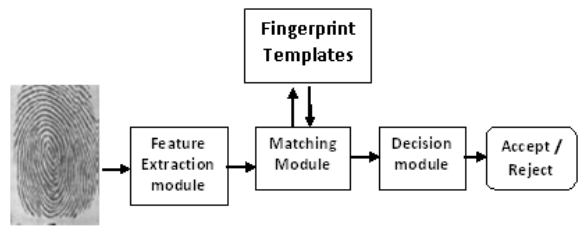
Figure 2: Overview of FPRS

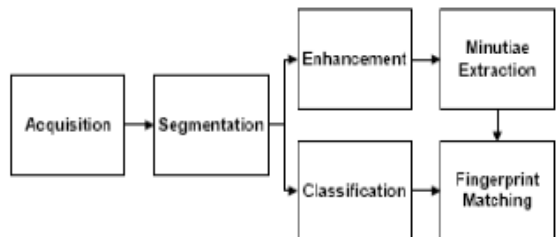
Figure 3: Overview of AFIS

*B. Fingerprint Segmentation*

Fingerprint segmentation is an important part of a fingerprint identification and verification system. This algorithm is based only on the block coherence of an image. Coherence gives us a measure of how well the gradients of the fingerprint image are pointing in the same direction[4]. In a window of size WxW around a pixel, the coherence is defined as Mc is the global mean of the coherence image and Sc is its global standard deviation. Holes in the segmentation mask are removed using morphological post processing. This mask on pixel-wise multiplication with the fingerprint image gives the segmented image.

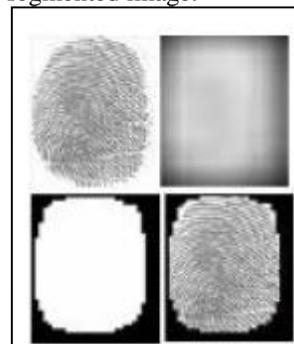
Figure 4: Segmented Image
A. Original Image, B. Coherence Image, C. Segmentation Mask, D) Segmented

*C. Fingerprint Enhancement*

The fingerprint enhancement algorithm mentioned in was found to be suitable for this application and was therefore used in the system. Better results are obtained using but it is slightly more time consuming. This algorithm calls for the development of a ridge frequency image IRF and ridge orientation IRO image for a fingerprint[5]. Gabor filters are used to enhance the fingerprint utilizing the ridge frequency and ridge orientation information Obtained from the frequency and orientation images obtained earlier the enhanced image IE is then binarized using adaptive thresholding to give a binarized image IEB. The binarized image is thinned to give IT. The thinned version is used for minutiae extraction.

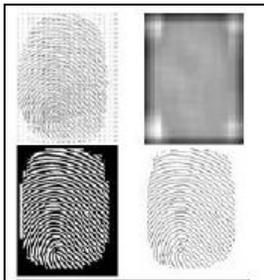

Figure 5: Minutiae Extraction A. Orientation Field, B. Ridge Frequency Image C. Enhanced Image D. Thinned Image

*D. Minutia Features*

The major Minutia features of fingerprint ridges are: ridge ending, bifurcation, and short ridge (or dot). The ridge ending is the point at which a ridge terminates. Bifurcations are points at which a single ridge splits into two ridges[6]. Short ridges (or dots) are ridges which are significantly shorter than the average ridge length on the fingerprint. Minutiae and patterns are very important in the analysis of fingerprints since no two fingers have been shown to be identical.

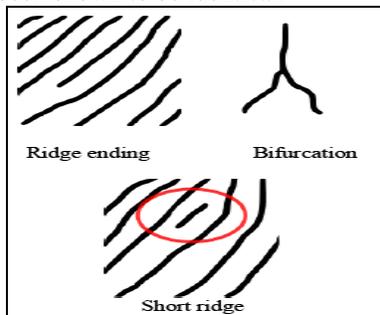

Figure 6: FP Images

*E. Minutiae Extraction*

Minutiae extraction was carried out using the crossing number approach. Crossing number of pixel "p" is defined as half the sum of the differences between pairs of adjacent pixels defining the 8-neighborhood of "p". Mathematically

$$cn(p) = \frac{1}{2} \sum_{i=1..8} |val(p_{i \mod 8}) - val(p_{i-1})|$$

Where *p0* to *p7* are the pixels belonging to an ordered sequence of pixels defining the 8-neighborhood of *p* and *val (p)* is the pixel value

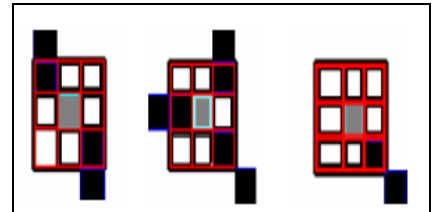

Figure 7: Non Minutiae Region

cn (p)=2, cn (p)=3 and cn (p)=1 representing a non minutiae region, a bifurcation and a ridge ending Crossing numbers 1 and 3 correspond to ridge endings and ridge bifurcations respectively[7]. An intermediate ridge point has a crossing number of 2. The minutiae obtained from this algorithm must be filtered to preserve only the true minutiae. The different types of false minutiae introduced during minutiae extraction include spike, bridge, hole, break, Spur, Ladder, and Misclassified Border areas. (See figure 8)

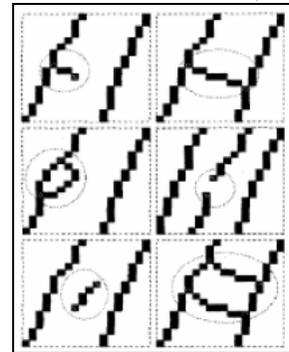

Figure 8: Different Types of False Minutiae A. Spike, B. Bridge, C. Hole, D. Break, E. Spur F. Ladder

*F. Minutiae Matching*

Let T and I be the representation of the template and input fingerprint, respectively[8]. Let the minutiae sets of the two Fingerprints be given by: A minutia $m_j$" in I and a minutia mi in T are considered to be matched if their spatial and orientation differences are within specified thresholds ro and θo.

T={$m_1, m_2, \ldots m_n$}  $m_i$={$x_i, y_i, \theta_i$}=1..m
I={m1, m2, \ldots mn}  $m_j$={$x_j, y_j, \theta_j$}=1..n

In this approach the minutiae sets are first registered using a derivative of the Hough transform, followed by fingerprint matching using spatial and orientation-based distance computation. The matching algorithm returns a percentage match score, which is then used to take the match-no match decision based on the security criterion.

## IV. IRIS RECOGNITION SYSTEM

The iris is the elastic, pigmented, connective tissue that controls the pupil. The iris is formed in early life in a process called morphogenesis[12]. Once fully formed, the texture is stable throughout life. It is the only internal human organ visible from the outside and is protected by the cornea. The iris of the eye has a unique pattern, from eye to eye and person to person.

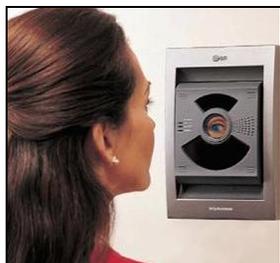
Figure 11 IRIS recognition system

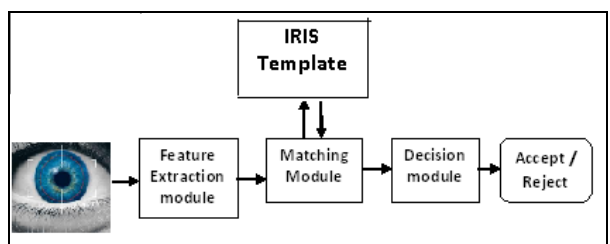
Figure 12 Overview of IRS

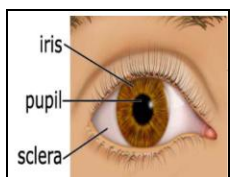
Figure13 Inner view of EYE

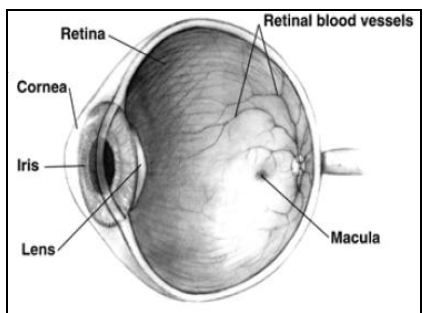
Figure14 Inner view of Retina

The iris image acquired from a 3CCD camera is localized by finding the center of pupil from the spectrum image. The radius of the pupil is the distance between the pupil center and nearest non-zero pixel[13]. The outer iris boundary is detected by drawing concentric circles of different radii from the pupil center and the intensities lying over the perimeter of the circle are summed up. Among the candidate iris circles, the circle having a maximum change in intensity with respect to the previous drawn circle is the outer iris boundary (shown in Figure 15). The annular region lying between pupil and iris boundary is transformed to polar co-ordinates to take into consideration the possibility of pupil dilation and appearing of different size in different images. From the normalized strip the eyelids are detected and removed.

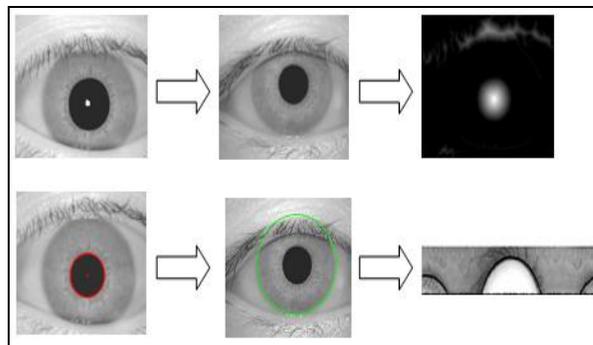
Figure 15 Steps involved in iris preprocessing and normalization

*A. Feature Extraction using Haar Wavelet and Circular Mellin operator*

*1.Haar Wavelet*

Haar wavelet is widely used in texture recognition algorithms .The input signal *S* (polarized iris image) is decomposed into approximation, vertical, horizontal and diagonal coefficients using the wavelet transformation and coefficients for the fourth and fifth levels are chosen to reduce space complexity and discard the redundant information. The iris code is generated by assigning one to the positive coefficient values and zero to negative values.

*2.Circular Mellin operators*

These "Circular Mellin" operators are invariant to both scale and orientation of the target and represent the spectral decomposition of the image scene in the polar-log coordinate system. Features in iris images are extracted based on the phase of convolution of polarized iris image with mellin operators. The iris code is one for positive phase values and zero for negative phase values.The iris codes generated using Haar Wavelet and Circular Mellin operators are matched using Hamming Distance approach.

*3.Combination of Haar Wavelet and Circular Mellin operator*

The individual matching scores generated by above mentioned classifiers are converted from distance to similarity score and are fused at matching score level for better performance of iris recognition.

VI. HOLISTIC FUSION

The purpose of this diagram is to illustrate the various levels of fusion for combining two (or more) biometric systems[14]. The three possible levels of fusion are: (a) fusion at the feature extraction level, (b) fusion at the matching score level, (c) fusion at the decision level.

(1) Fusion at the feature extraction level: The data obtained from each sensor is used to compute a feature vector. As the features extracted from one biometric trait are independent of those extracted from the other, it is reasonable to concatenate the two vectors into a single new vector. The new feature vector now has a higher Dimensionality and represents a persons identity in a different (and hopefully more discriminating) hyperspace. Feature reduction techniques may be employed to extract useful features from the larger set of features.

(2) Fusion at the matching scores level: Each system provides a matching score indicating the proximity of the feature vector with the template vector[15]. These scores can be combined to assert the veracity of the claimed identity Techniques such as logistic regression may be used to combine the scores reported by the two sensors. These techniques attempt to minimize the FRR for a given FAR (Jain et al., 1999b).

(3) Fusion at the decision level: Each sensor can capture multiple biometric data and the resulting feature vectors individually classified into the two classes—accept or reject. A majority vote scheme, such as that employed in (Zuev and Ivanon, 1996) can be used to make the final decision.

The different biometrics systems can be integrated at multi-classifier and multi-modality level to improve the performance of the verification system. However, it can be thought as a conventional fusion problem i.e. can be thought to combine evidence provided by different biometrics to improve the overall decision accuracy. The multimodal biometric system is developed at multi-classifier and multi-modalities level. At multi- classifier level, multiple algorithms are developed and combined for traits like fingerprint and iris. The following steps are performed for fusion at classifier level:

S1: Given a query image as input, features are extracted by the individual recognizers and then an individual comparison algorithm for each recognizer compares the set of features and calculates the matching scores or distances corresponding to each recognizer for various traits.
S2: The scores/distances obtained in S1 are normalized to a common range between 0 to 1.
S3: These scores are then converted from distance to similarity score by subtraction from 1 if it is a dissimilarity score. For example the dissimilarity scores, in case of fingerprint recognition using reference point algorithm ($D_{Ref}$), iris recognition using Haar Wavelet ($D_{Haar}$) and Circular Mellin operator ($D_{Mellin}$) are converted to similarity scores ($MS_{Ref}$, $MS_{Haar}$, $MS_{Mellin}$)
S4: The matching scores are further rescaled so that threshold value becomes same for each recognizer.
S5: Then the combined matching score is calculated by fusion of the matching scores of multiple classifiers using sum rule technique.

$$MS_{Finger} = \frac{\alpha \times MS_{Ref} + \beta \times MS_{MIN}}{2}$$

$$MS_{Iris} = \frac{\alpha \times MS_{Haar} + \beta \times MS_{Mellin}}{2}$$

where α and β are the weights assigned to individual classifiers. Currently equal weightage is given to each classifiers and the value of α and β is one. The multimodal biometric system at IITK is developed by integrating four traits i.e., fingerprint and iris at matching score level. Based on the proximity of feature vector and template, each subsystem computes its own matching score. These individual scores are finally combined into a total score, which is passed to the decision module. The same steps for fusion at classifiers level are followed for multiple modalities level i.e., matching scores are computed for each trait (fingerprint and iris) followed by normalization to the common scale and distance to similarity score conversion for all the four traits. The matching scores are further rescaled so that the threshold value becomes common for all the subsystems. Finally, the sum of score technique is applied for combining the matching scores of two traits i.e., fingerprint and iris. Thus the final score $MS_{Final}$ is given by,

$$MS_{Final} = \frac{1}{4}(a \times MS_{Finger} + b \times M_{iris}))$$

where $MS_{Finger}$ = matching score of fingerprint, $MS_{Iris}$ = matching score of iris, and a, b, c and d are the weights assigned to the various traits. Currently, equal weightage is assigned to each trait so the value of a, b, c and d is one. The final matching score ($MS_{Final}$) is compared against a certain threshold value to recognize the person as genuine or an imposter.

## VI. CONCLUSION

Multimodal biometric door lock system is a biometric authentication technology for controlling door access in a convenient way by applying the high level of security provided by fingerprint and iris. Biometrics has started to be applied for civilian-identification purposes like passport inspection. We consider that fingerprint and iris can take a leading role in such applications; accordingly, we will strive to push forward commercialization and development of this product to make Multimodal biometric door lock authentication more convenient.

VII.REFERENCES

[1] S. Pravinthraja and Dr.K. Umamaheswari "Multimodal Biometrics for Improving Automatic Teller Machine Security S, Bonfring International Journal of Advances in Image Processing, Vol. 1, Special Issue, December 2011.
[2] A. Ross & A. K. Jain, *Information Fusion in Biometrics*, Pattern Recognition Letters, 24 (13), pp. 2115-2125, 2003.
[3] A.S. Tolba & A. A. Rezq, *Combined Classifier for Invariant Face Recognition*, Pattern Analysis and Applications, 3(4), pp. 289-302, 2000
[4] A. Ross, A. K. Jain & J.A. Riesman, *Hybrid fingerprint matcher*, Pattern Recognition, 36, pp. 1661–1673, 2003
[5] W. Yunhong, T. Tan & A. K. Jain, *Combining Face and Iris Biometrics for Identity Verification*, Proceedings of Fourth International Conference on AVBPA, pp. 805-813, 2003
[6] S. C. Dass, K. Nandakumar & A. K. Jain, *A principal approach to score level fusion in Multimodal Biometrics System*, Proceedings of ABVPA, 2005
[7] J. Kittler, M. Hatef, R. P. W. Duin, & J. Mates, *On combining classifiers*, IEEE Transactions on Pattern Analysis and Machine Intelligence, 20(3), pp. 226–239, 1998
[8] G. Feng, K. Dong, D. Hu & D. Zhang, *When Faces Are Combined with Palmprints*: *A Novel Biometric Fusion Strategy*, ICBA, pp. 701-707, 2004
[9] I. Craw, D. Tock & A. Bennett, *Finding Face Features*, Proceedings Second European Conference Computer Vision, pp. 92-96, 1992.
[10] C. Lin & Kuo-Chin Fan, *Triangle-based approach to the detection of human face*, Pattern Recognition, 34, pp. 1271-1284, 2001.
[11] Juwei Lu, K. N. Plataniotis & A. N. Venetsanopoulos, *Face Recognition Using Kernel Direct Discriminant Analysis Algorithms*, IEEE Transactions on Neural Networks, 14 (1), pp. 117-126, 2003.
[12] D. S. Bolme, J. R. Beveridge, M. L. Teixeira & B. A. Draper, *The CSU Face Identification Evaluation System: Its Purpose, Features and Structure*, Proceedings 3rd International Conference on Computer Vision Systems, 2003.
[13] N.K.Ratha, K.Karu, S.Chen & A.K.Jain, *A Real-time Matching System for Large Fingerprint Database*, IEEE Transactions on Pattern Analysis and Machine Intelligence, 18 (8), pp. 799-813, 1996
[14] A.K.Jain, L.Hong & R.M.Bolle, *On-line Fingerprint Verification*, IEEE Transactions on Pattern Analysis and Machine Intelligence, 19(4), pp. 302-313, 1997
[15] L. Hong, Y. Wan & A. Jain, *Fingerprint Image Enhancement: Algorithm and Performance Evaluation*, IEEE Transcations on Pattern Analysis and Machine Intelligence, 20(8), pp. 777-789, 1998